\documentclass{llncs}

\usepackage{algorithmicx}
\usepackage{algorithm}
\usepackage{algpseudocode}
\usepackage{pifont}
\usepackage{cite}
\usepackage{csquotes}
\usepackage{graphicx}
\usepackage{hyperref}
\usepackage{url}
\usepackage{comment}
\usepackage{booktabs}

\graphicspath{ {figures/} }

\usepackage{todonotes}
\usepackage{cleveref}

\usepackage{color}

\title{Talking Open Data}
\author{Sebastian Neumaier \and Vadim Savenkov \and Svitlana Vakulenko}
\institute{Vienna University of Economics and Business \\ Institute for Information Business \\ \email{\{firstname.lastname\}@wu.ac.at}}

\date{14 March 2017}

\begin{document}

\maketitle

\begin{abstract}
% TODO vertical search engine
Enticing users into exploring Open Data remains an important challenge for the whole Open Data paradigm. Standard stock interfaces often used by Open Data portals are anything but inspiring even for tech-savvy users, let alone those without an articulated interest in data science. To address a broader range of citizens, we designed an open data search interface supporting natural language interactions via popular platforms like Facebook and Skype. Our data-aware chatbot answers search requests and suggests relevant open datasets, bringing fun factor and a potential of viral dissemination into Open Data exploration. The current system prototype is available for Facebook\footnote{\url{https://m.me/OpenDataAssistant}} and Skype\footnote{\url{https://join.skype.com/bot/6db830ca-b365-44c4-9f4d-d423f728e741}} users.
\end{abstract}

% demo submissions are at most five pages long
% Accepted posters and demos will be included into the “minute madness” pitch

% The submissions must clearly demonstrate relevance to the Semantic Web and the topics of interest of ESWC 2017. Decisions about acceptance will be based on relevance to the Semantic Web, originality, potential significance, topicality, and clarity.

\section{Introduction}

% Motivation
The European Commission defines Open Data portals as ``web-based interfaces designed to make it easier to find re-usable information''\footnote{\url{https://ec.europa.eu/digital-single-market/en/open-data-portals}}. It is exactly the task of finding re-usable information, however, where current data portals fall short: they focus on supporting users to find files and not the information. It remains a tedious task for users to drill out and understand the information behind the data. According to the EU Data Portal study, 73\% of the open data users characterize finding data as \emph{Difficult} or \emph{Very Difficult}\footnote{\url{https://www.europeandataportal.eu/en/highlights/barriers-working-open-data}}.

Currently, the data is brought to end users not directly but through apps, each focusing on a specific service and based on a handful of open datasets. The app ecosystem, promoted by Open Data portals, is thriving. The downside of this approach, however, is that the data remains hidden from the users, who have to rely on IT professionals to get insights into it.

An orthogonal approach is to lower the entry barrier for working with open data for a broader audience. Being a moving target, as powerful data analysis methods become increasingly complicated, this approach still has a number of unique benefits. One can argue that it is embedded in the spirit of the Open Data movement itself, to enable and empower citizens to analyze the data, to be able to draw and share their own conclusions from it.
Working with raw data is and will probably remain challenging for non-experts, and thus the easier it is to find the right dataset and to understand its structure, the more energy a user has to actually work with it. Despite (or rather due to) the growing number of Open Data portals and ever increasing volumes of data served through them, their accessibility for non-technicians is still hampered by the lack of comprehensive and intuitive deep search, and means of integrating data across domains, languages and portals.

This demo showcases a novel natural-language interface, allowing users to search for open datasets by talking to the chatbot on a social network. We also address the challenge of cross-lingual dataset search that goes beyond the monolingual prototype\footnote{\url{https://m.me/OpenDataATAssistant/}} we developed earlier for two Austrian open data portals. To improve user experience, it embeds a state-of-the-art approach to semantic linking for natural language texts into a dialogue-based user interface.

Our hypothesis is that using a popular and convenient communication channel opens new possibilities for interactive search sessions. The inherent interactivity of a chat session makes it easy to enhance user experience with context-based and personalized elements. We envision our prototype to be the first step towards an intelligent dialogue system supporting contextual multilingual semantic search~\cite{Baeza-Yates:2017:TYW:3018661.3022744}, focused on retrieval of datasets as well as the individual data items from them. In the future work we plan to extend the chatbot to search within the content of the datasets rather than merely in the metadata.

% \todo{Why semantic web (integration)?}

\section{Chatbot architecture}

% ANNOTATION OF DATASETS
We implemented a prototype as a proof-of-concept by pooling and annotating 18k datasets from seven Open Data portals with dataset descriptions in seven different languages (see Table~\ref{tab:portals}).
The front-end is designed using Microsoft Bot Framework\footnote{\url{https://dev.botframework.com}}, which connects the implementation to both Facebook Messenger and Skype platforms.

\begin{table}[]
 \vspace{-0.3cm}
\caption{List of selected data portals and respective languages\label{tab:portals}}
\centering
\begin{tabular}{l|l| c | r }
\toprule
Portal & Country & Language & Datasets \\
\midrule
dati.trentino.it	& Italy     & IT & 5285 \\
data.gov.ie	        & Ireland   & EN & 4796 \\
datamx.io		    & Mexico    & SP & 2767 \\
data.gv.at	        & Austria   & DE & 2323 \\
dados.gov.br	    & Brazil	& PT & 2061 \\
beta.avoindata.fi	& Finland   & FI & 820 \\
www.nosdonnees.fr	& France    & FR & 290 \\
\bottomrule
\end{tabular}
 \vspace{-0.5cm}
\end{table}

\begin{figure}[t]
% \frame{\includegraphics[width=\textwidth]{figures/arch_dia_oda.pdf}}
\includegraphics[width=\textwidth]{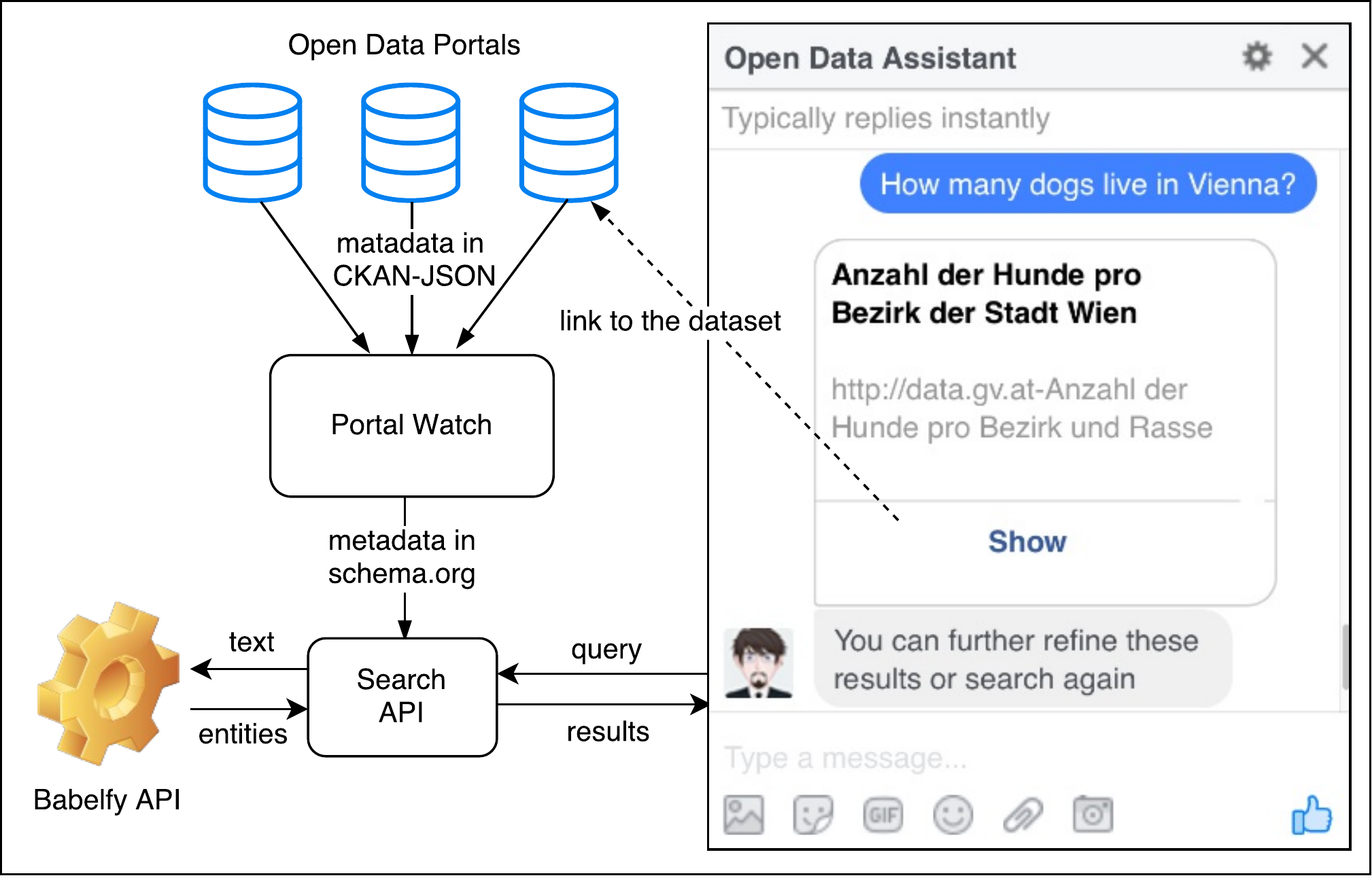}
\caption{Open Data Assistant chatbot. The system integrates metadata from different open data portals into a unified Schema.org format and enriches it with the concepts extracted from text via the Babelfly API. The chatbot interface provides access to the semantic (cross-lingual) open data Search API over the dataset metadata.\label{fig:architecture}}
% \caption{Screenshot and the system architecture.}
 \vspace{-0.3cm}
\end{figure}

%\subsection{Linking \& annotation of dataset descriptions}
\textbf{Collection and annotation of datasets.} 
The search results of our chatbot application are based on enrichment of dataset descriptions with BabelNet synsets~\cite{Moroetal:14tacl}. 
Initially, we use the Open Data Portal Watch (ODPW) framework\cite{Neumaier17} to collect the dataset descriptions of the selected portals. ODPW harvests the metadata descriptions and maps them to the Schema.org standard vocabulary (cf. \cref{fig:architecture}).
We extract title, natural language description and keywords from these metadata and identify their language using langdetect\footnote{\url{https://pypi.python.org/pypi/langdetect}} Python package.
Then, we provide them as a single concatenated string (title, description, and keywords) for each dataset alongside the detected language to the Babelfy API\footnote{\url{https://babelfy.org/guide}} to detect and disambiguate entities and concepts within this string. The Babelfy API provides a list of corresponding ``babelSynsetIDs'' for an input string, language-independent entity identifiers in the BabelNet framework.

To deliver a good performance for the search functionality, we built an Elasticsearch index from the Schema.org dataset descriptions and the corresponding BabelNet entities. This allows us to retrieve all dataset descriptions that are annotated by a specific BabelNet entities and aggregate over the top co-occurring entities.

\textbf{Search API interactions.}
There are two modes of interaction and obtaining search results in the chatbot interface. First, the user can issue a free text search query. Our search API hands this input over to the Babelfy API which provides a list of disambiguated concepts and entities. 
%We query and return all datasets that hold \textit{any} of these input IDs to the user interface and rank them by the number of the IDs for each dataset. 
We query our Elasticsearch index for \textit{any} of these entities and return all matching datasets.
The datasets get ranked by the number of matching entities. For instance, if a query includes the entites \textit{dog} and \textit{vienna}, then a dataset annotated with both of these gets ranked higher than a dataset annotated with either \textit{dog} or \textit{vienna}.

% Including and querying the ``real'' input text as a second ranking criterion would be a straightforward improvement and would result in a better ranking in many cases (e.g., if the dataset is not annotated properly). However, we deliberately want to test and focus on a pure semantic-based search for our prototype.

In the second interaction step the user can refine the search results by selecting one of the top co-occurring concepts and entities. We then use the selected entities to filter the result set, i.e., the selected entities \textit{must} occur in the dataset description. This way we implicitly implement both AND/OR query operators. 

%In the back-end we use Babelfy API\footnote{\url{http://babelfy.org/guide}}~\cite{Moroetal:14tacl} for entity disambiguation, which helps us to link datasets across languages based on shared entities.

\section{Usability study}
% \textbf{Participants}
Seven participants took part in a usability study designed to evaluate our system prototype. We asked the participants to complete a predefined search task and reflect on their experience of using the system. The search task was to find the official statistics data from different countries concerning climate change so that the participants could also experiment using various keywords related to the topic of climate change, e.g. air temperatures, snow level, etc.

% All participants are of diverse backgrounds: with and without professional computer science education.

% \textbf{Task description.} 

% \section{Results}

Most of the participants found the system useful but in some cases limited in scope and functionality. Suggestions from the users include: (1) complementing open data with additional resources, such as Wikipedia; (2) user-specific answers, such as adjusting the language of interaction and geolocation-relevant queries; (3) context-specific answers, i.e. ability to follow up and refine the previous query. Often the participants were not able to assess the quality of the produced results when they were in an unfamiliar to the user language. More details on the evaluation task and results are available on-line\footnote{\url{https://github.com/vendi12/oda_evaluation}}.

% TODO Interaction and context are important for understanding user intent. Query refinement techniques such as query expansion, query suggestion, relevance feedback improve ranking.

% It allowed them to get access to various open data resources in different languages.

% Since our implementation is based on Babelfy API, it also inherits its limitations. For example, Babelfy disambiguation algorithm is case-sensitive, e.g. query for ``Vienna'' keyword returns many results, while ``vienna'' returns zero matches.

% For details on the please refer to our case study at https://github.com/WDAqua/Pipeline

% \section{Discussion and future work}

% \todo{describe ranking score limitation distinguish between different senses help from user request to disambiguate query}
% * table (column) semantics, linking similar tables (columns)~\cite{NeumaierUPP16}
% * question answering from tables~\cite{DBLP:conf/ijcai/YinLLK16}

% and its implementation.

% Sample queries

% e.g. http://crosscloud.org/2016/www-mansour-pdf.pdf

\section{Related Work}

The originality of our system is in applying chatbot interface to the dataset search, which, as we believe based on our early evaluation, has large potential for popularizing and promoting open data, e.g. through easy access and gamification component. 

Chatbots, e.g. Google Allo\footnote{\url{https://allo.google.com}}, recently gained an increased attention in the developers' communities worldwide. They integrate cutting-edge technologies, such as auto-reply, image and speech recognition, enhancing them with multimedia elements, which results in an attractive interface accessible also for users without IT background. Chatbot UI arguably provides a more natural and personalized way of human-computer interaction, as opposed to the traditional ``book-like'' web page. To the best of our knowledge ours is the first chatbot focusing on dataset search. 

The LingHub data portal\footnote{\url{http://linghub.lider-project.eu}}~\cite{DBLP:conf/i-semantics/McCraeC15} is another example of cross-lingual data search implementation. It integrates language resources from Metashare, CLARIN, etc. using RDF, DCAT and SPARQL on the metadata level. Similar to our system, LingHub employs the Babelfy disambiguation algorithm.

\section{Conclusion}

% \todo{Wrap-up}
We present a prototype of a conversational agent for Open Data search. The early user evaluation showed that such a cross-lingual dialog-based system has the potential to enable an easier access to Open Data resources. 

% integration of tabular data
The set of indexed portals can be easily extended since we rely on the ODPW framework that provides mapping of metadata from over 260 data portals into a homogenized schema, and Elasticsearch, which implements scalable search functionality. We aslo plan to extend the chatbot to search within the content of the datasets rather than merely in the metadata. %For instance, if we label a column of cities by their named entities and extract the common class ``city'' we allow users to search not only by instances but also classes of entities.
Furthermore, the user query understanding needs to be enhanced to improve the results ranking. One way to facilitate it in the interactive chat context would be through asking user the questions to disambiguate the query, e.g. ``Did you mean apple as a fruit or as a company Apple Inc.?''
%Currently, we collect all possible concepts in the input query, however, this might lead to unexpected or unwanted results (e.g., if users ask for ``datasets describing climate change'' we also extract the entity ``dataset'' and use it in our search query). 
% In \cite{RicoUC15}, patterns are used to extract instances and properties in natural language queries and translate them to SPARQL queries.
% cimiano http://ceur-ws.org/Vol-1472/IESD_2015_paper_12.pdf
% One of the benefits of a dialogue-based user interface is the user interaction and the availability of a personal query history. 

% take this one step further by asking the users to disambiguate certain terms in case of ambiguous input and by providing search refinement options such as categories or location-based information of the datasets.

% Question answering

% \section{Demonstration plan}
\paragraph{\bf Demonstration plan.}
% \todo{textual description of the demonstration to be given at the conference}

Conference participants will be able to interact with the chatbot via Facebook and Skype. They will be free to experiment and come up with their own queries to the system. We will also provide the participants with the query samples that showcase both the strengths and the pitfalls of the current approach.

% \section*{Acknowledgements}
\paragraph{\bf Acknowledgements.}
This work was supported by the Austrian Research Promotion Agency (FFG) under the projects ADEQUATe (grant no. 849982) and CommuniData (grant no. 855407).

\bibliographystyle{splncs03}
\bibliography{refs}

\begin{thebibliography}{1}
\providecommand{\url}[1]{\texttt{#1}}
\providecommand{\urlprefix}{URL }

\bibitem{Baeza-Yates:2017:TYW:3018661.3022744}
Baeza-Yates, R.: Ten years of wisdom. In: Proceedings of the Tenth {ACM}
  International Conference on Web Search and Data Mining, {WSDM} 2017,
  Cambridge, United Kingdom, February 6-10, 2017. pp. 1--2 (2017)

\bibitem{DBLP:conf/i-semantics/McCraeC15}
McCrae, J.P., Cimiano, P.: Linghub: a linked data based portal supporting the
  discovery of language resources. In: Joint Proceedings of the Posters and
  Demos Track and 1st Workshop on Data Science: Methods, Technology and
  Applications (DSci15) of 11th International Conference on Semantic Systems -
  SEMANTiCS 2015, Vienna, Austria, September 15-17, 2015. pp. 88--91 (2015)

\bibitem{Moroetal:14tacl}
Moro, A., Raganato, A., Navigli, R.: {Entity Linking meets Word Sense
  Disambiguation: a Unified Approach}. Transactions of the Association for
  Computational Linguistics (TACL)  2,  231--244 (2014)

\bibitem{Neumaier17}
Neumaier, S., Umbrich, J., Polleres, A.: Lifting data portals to the web of
  data. In: WWW2017 Workshop on Linked Data on the Web (LDOW2017), Perth,
  Australia, April 3-7, 2017 (2017)

\end{thebibliography}

\end{document}